\def \etal         {{\it et~al.} }
\def \flam         {\hbox{ergs s$^{-1}$ cm$^{-2}$~\AA $^{-1}$}}
\def \kms          {\rm{\hbox{km s$^{-1}$}}}

\def \Lya          {\hbox{Ly$\alpha$}}

\def \zaz          {{$z_a\kern -1.5pt \approx\kern -1.5pt z_e$}}
\def \zllz         {{$z_a\kern -3pt \ll\kern -3pt z_e$}}

\def \zgz          {{\kiA z\lower 3pt \hbox{a} $>$ z\lower 3pt \hbox{e}\ }}
\documentstyle[11pt,paspconf]{article}

\begin{document}

\title{High-Velocity Narrow-Line Absorbers in QSOs}
\author{F. Hamann, T. Barlow, R.D. Cohen, 
V. Junkkarinen \& E.M. Burbidge}
\affil{Center for Astrophysics \& Space Sciences, University of California 
-- San Diego, La Jolla, CA, 92093-0424}

\begin{abstract}
We discuss the identification and kinematics of intrinsic 
narrow-line absorbers in three QSOs. The line-of-sight 
ejection velocities in the confirmed intrinsic systems range 
from $\sim$1500~\kms\ to $\sim$51,000~\kms . The ratio 
of ejection velocity to line width, $V/\Delta V$, in these 
systems ranges from $\sim$3 to $\sim$60, in marked contrast to 
the BALs where typically $V/\Delta V\sim 1$. We speculate on the 
possible relationship between the BAL and narrow-line absorbers. 
Like the BALs, intrinsic narrow-line systems appear to be rare, but 
the outflows that cause them might be common 
if they cover a small fraction of the sky as seen 
from the QSO.
\end{abstract}


\section{Introduction}

We are involved in a program to identify and study the intrinsic 
absorption-line systems in QSOs using spectra from 
the {\it Hubble Space Telescope} ({\it HST}) and the Keck and Lick 
Observatories. The well-known broad absorption lines (BALs) 
detected in roughly 10--15\% of QSO spectra clearly form in 
high-velocity outflows from the QSOs. However, the location of  
narrow-line absorbers in QSO spectra must be examined on a 
case-by-case basis. Narrow-line systems 
could form (1) in ejecta or infall very near the QSO engine, 
(2) in the extended host galaxy of the QSO, (3) in a nearby cluster 
galaxy, or (4) in cosmologically intervening gas. We will refer to 
the category (1) systems (which include the BALs) as intrinsic and 
all others as intervening. In this proceedings we discuss 
the identification and kinematics of intrinsic narrow-line 
absorbers in three radio-quiet QSOs: PG~0935+417 ($z_e = 1.966$), 
UM~675 ($z_e = 2.150$) and Q~2343+125 ($z_e = 2.515$). 
Some results on the ionization and metal abundances 
in these and other intrinsic systems are presented in 
our accompanying paper (by Hamann {\it et al.}) in this volume.

\section{Observational Results}

\begin{figure}
\plotfiddle{fhamann1.ps1}{2.7in}{0.0}{43.}{43.}{-165.0}{-34.0}
\caption{Lick-KAST spectra of PG~0935+417. The 
\zaz\ lines are labeled above and the high-velocity C~{\sc iv} feature 
is labeled below. The 1996 data are scaled to match the 1993 flux 
(in $10^{-15}$ \flam ) at 4200--4400~\AA .} 

\plotfiddle{fhamann1.ps2}{3.49in}{0.0}{43.}{43.}{-120.0}{-54.0}
\caption{Keck-HIRES spectra of PG~0935+417 (from Feb. 1996) 
showing C~{\sc iv} lines in the \zaz\ (top panel) 
and high-velocity systems (lower) on velocity scales relative to 
the emission redshift for 1548.20~\AA . 
The lower panel also shows the 1993 
Lick data for comparison.} 
\end{figure}

Figure 1 compares two spectra of PG~0935+417 obtained at Lick in 
1993 and 1996 with roughly 220~\kms\ resolution. Both observations 
show strong \zaz\ lines plus a relatively broad absorption feature 
near 3870~\AA . The broad feature varied significantly between the 
two measurements. We attribute the $\sim$3870~\AA\ feature to extremely 
high-velocity C~{\sc iv} based on probable detections of corresponding 
N~{\sc v} and O~{\sc vi} absorption in {\it HST} spectra (not shown) 
and because an identification with blueshifted Si~{\sc iv} 
is ruled out by the absence of C~{\sc iv} absorption at $\sim$4300~\AA . 
Figure 2 shows Keck observatory spectra at $\sim$7~\kms\ resolution of 
C~{\sc iv} in the \zaz\ and broad high-velocity systems. The Keck data 
(from Feb. 1996) show that the \zaz\ absorber (upper panel in Fig. 2) can be 
regarded 
as two systems with dramatically different line profiles and at least three 
components in each system. The high-velocity C~{\sc iv} absorber (lower 
panel) is shifted  $\sim$51,000~\kms\ from the emission redshift and 
maintains a smooth profile at high resolution. (The high-velocity feature 
spans more than one Keck-HIRES echelle order and its profile is forced to 
match the March 1996 Lick data in this plot.) 

Figure 3 shows Keck and other spectra of intrinsic lines in 
UM~675 and Q~2343+125 (see also Hamann \etal 1995, 1997a and 1997b). 
Both systems are dominated by relatively broad components 
(FWHM~$\sim$~450~\kms ) that varied considerably between observations 
and are well-resolved at the 7--9~\kms\ resolution of these Keck data. 
Much narrower lines (labeled in the figure) are also present in both 
systems. For UM~675, the figure compares Keck-HIRES spectra of several 
\zaz\ lines measured in September 
1994 to observations by Sargent, Boksenberg \& Steidel (1988) obtained 
at Palomar observatory in November 1981. For Q~2343+125, 
the figure compares two Keck observations (September 1994 and 
October 1995) of high-velocity C~{\sc iv} lines with data obtained 
at Palomar by Sargent \etal (1988) in October 1984.  
N~{\sc v}, Si~{\sc iv} and possibly \Lya\ are also detected in this 
system (see Hamann \etal 1997b). The absence of narrow lines in the 
ratio of the two Keck spectra of Q~2343+125 shows that these 
components did not vary. 

\begin{figure}
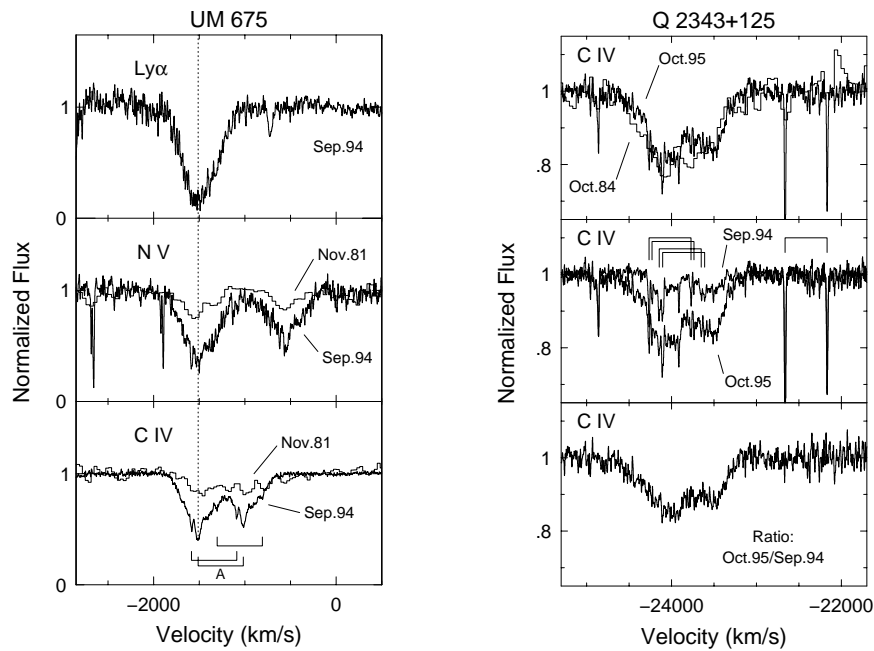

\plotfiddle{fhamann1.ps3}{2.89in}{90.0}{40.}{40.}{20.0}{-87.0}
\plotfiddle{fhamann1.ps4}{0.0in}{0.0}{40.}{40.}{-42.0}{-63.0}
\caption{Multi-epoch observations of intrinsic absorption lines in UM~675 
(left panels) and Q~2343+125 (right) on velocity scales relative to the 
emission redshifts (appropriate for 1548.20 \AA\ in C~{\sc iv}, etc.).} 
\end{figure}

\section{Evidence for Intrinsic Absorption}

Barlow \& Sargent (1997) and Hamann \etal (1997a) recently summarized 
observational signatures that can distinguish intrinsic from 
intervening absorption-line systems (also Barlow \etal this 
volume). These signatures include (1) time-variable line strengths, 
(2) partial line-of-sight coverage 
of the background light source(s), and (3) absorption-line profiles 
that are broader and smoother than intervening systems when 
measured at high spectral resolutions (cf. Blades 1988). 
The line-of-sight coverage fraction can be derived from 
multiplet ratios whose relative optical depths (proportional to 
$f\lambda$) are known from atomic physics. Partial coverage 
weakens the absorption troughs and leads to 
multiplet ratios closer to unity than expected from the measured 
depths of the lines. For doublets with $f\lambda$ ratios 
of $\sim$2, such as C~{\sc iv} and N~{\sc v}, the coverage 
fraction, $C_f$ ($0\leq C_f\leq 1$), at any velocity in the profiles 
is given by, 
\begin{equation}
C_f = {{I_1^2 - 2I_1 + 1}\over{I_2 - 2I_1 +1}} 
\end{equation} 
where $I_1$ and $I_2$ are the residual intensities in the weaker 
and stronger doublet lines, respectively, normalized by the continuum 
intensity (also Hamann \etal 1997a; Barlow \& Sargent 1997). 
Note that the coverage fraction includes both 
direct and scattered sources of radiation; values of $C_f < 1$ 
might result from complete coverage along direct lines-of-sight 
but zero coverage of the scattered flux 
(see papers on polarization in this volume).  

The dominant broad line components in UM~675 and Q~2343+125 
are clearly intrinsic because they meet all three criteria above. 
Their time-variability 
and broad and smooth profiles are evident from Figure 3. We 
estimate coverage fractions at the center of the broad C~{\sc iv} 
profiles of $\sim$50\% in UM~675 and $\la$20\% in Q~2343+125 
(also Hamann \etal 1995, Hamann \etal 1997a and 1997b).
The status of the much narrower lines 
in these QSOs is unknown. In Q~2343+125, the fact that the 
narrow C~{\sc iv} lines did not vary with the broad components indicates 
that the narrow features form in a physically distinct region. 
In UM~675, the narrow systems 
blended with the broad lines are probably intrinsic and related 
to the broad-component gas because (1) their profiles are still 
broader than typical intervening systems and (2) the doublet ratio 
in the strongest narrow-line system (labeled `A' in Figure 3) indicates 
partial coverage ($C_f < 50$\%). 

We also identify 2 of the 3 systems mentioned above for PG~0935+417 
(Figs. 1 and 2) as intrinsic. The high-velocity 
C~{\sc iv} feature (at $\sim$51,000~\kms ) is time-variable and has a 
much broader and smoother profile than ``known'' intervening absorbers. 
We cannot estimate the coverage fraction for this system because the 
absorption feature is much wider than the doublet separation. 
The \zaz\ system at roughly $-$2800~\kms\ 
is also intrinsic based on its relatively broad profiles, partial 
line-of-sight coverage, and a possible change in the absorption 
depth near $\sim$3000~\kms\ (from a second Keck-HIRES spectrum 
not shown). We estimate $C_f \la 75$\% near the line centers of 
the 3 components in this system. The origin of the narrower \zaz\ 
lines at roughly $-$1600~\kms\ in PG~0935+417 is uncertain; 
they showed no evidence for variability, their coverage fractions 
are consistent with unity, and the narrow profiles are consistent 
with intervening absorption. 

\section{Outflow Kinematics and Morphologies}

The line-of-sight ejection velocities in the systems discussed here 
range from roughly 1500~\kms\ in UM~675 to 51,000~\kms\ in 
PG~0935+417. These outflow velocities span the same range as BALs 
measured in other spectra, but in BALs the ratio of line 
centroid velocity to line width $V/\Delta V$ is typically of order 
unity (Weymann \etal 1991). Here we find  $V/\Delta V$ ranging 
from $\sim$3 in UM~675 to $\sim$60 in Q~2343+125. The system at 
$-$51,000~\kms\ in PG~0935+417 has $V/\Delta V\sim 34$. These large 
$V/\Delta V$ ratios require either (1) a highly-collimated steady 
flow that crosses our line-of-sight to the emission source(s) 
and thereby reveals only part of its full velocity extent, or (2) 
discrete ``blobs'' that might arise, for example, from episodic 
ejection events. The line widths, while much narrower than BALs, are still 
many times larger than the thermal speeds ($\la$5~\kms\ for carbon at a 
temperature of $\la$20,000~K). Therefore the absorbers have a range of 
non-thermal line-of-sight velocities and they cannot be single, 
coherently moving clouds. For any geometry where the absorbers are  
moving along our line-of-sight to the continuum source(s), the 
finite line widths imply that the absorbing regions are radially 
expanding or contracting or have some other large internal 
motion such as turbulence. 

Whatever flow geometry obtains, the measured ions (from H~{\sc i} and 
Si~{\sc iv} to C~{\sc iv}, N{\sc v} and O~{\sc vi} in different 
sources) always appear 
at the same velocity (within the uncertainties). Evidently, the 
ionization structures are not highly stratified in velocity. 
Another constraint is the lack of acceleration in the Q~2343+125 
absorber. Figure 3 shows that this system shifted by 
$\la$100~\kms\ in 3.1 yrs in the QSO rest frame (11 years observed). 
In contrast, the centroid of the extreme high-velocity absorber 
in PG~0935+417 shifted by roughly +900~\kms\ in 1 yr in the QSO rest frame. 
However, the complex variation in that absorption profile suggests 
that the apparent shift was caused by changes in the optical 
depth structure rather than acceleration of the gas.

\section{Discussion}

We have not (yet) conducted a systematic survey for intrinsic narrow-line 
absorbers in QSOs, but one is clearly needed to understand the frequency 
of these systems, the range of their kinematics and physical conditions,  
and their possible relationship to the BALs and other QSO properties 
such as radio loudness and soft X-ray absorption. It is important to keep 
in mind that narrow metal-line systems that do not show any of the 
signatures of intrinsic absorption (\S 3) might be intrinsic nonetheless. 
It is our experience from inspection of moderate-resolution absorption-line 
survey spectra, such as Sargent \etal (1988), that high-velocity systems 
with line widths above a few hundred \kms\ (as in PG~0935+417 
and Q~2343+125) do not occur in more than a few percent of QSOs. 
Nonetheless, these absorbers might be ubiquitous in QSOs if, 
like the BAL gas, they cover only a small fraction of the sky as seen 
from the central QSO. It is also our experience, based on PG~0935+417, 
Q~2343+125, and one other QSO not shown here (Q~0151+048), that 
the broader high-velocity systems are extremely variable; we detected 
substantial changes in our first two observations of each of these 
systems. Jannuzi \etal (1996) reported {\it HST} observations of 
a system that appears similar to the high-velocity absorber in 
PG~0935+417 -- at roughly $-$56,000~\kms\ in the low-redshift 
QSO PG~2302+029. We expect that follow-up observations will 
show line-strength changes and confirm the intrinsic nature of 
that system. Since velocities above 50,000~\kms\ are extremely 
rare among BALs, the existing data suggest that narrow-line intrinsic 
absorbers can reach these speeds much more readily. 

Many narrow-line \zaz\ systems are known in radio-loud quasars, 
but the three QSOs discussed here, plus Q~0151+048 and PG~2302+029 
mentioned above, are all radio-quiet. Perhaps it is only systems 
with high velocities and/or relatively broad and smooth line 
profiles (i.e. ``mini-BALs'') that are like the BALs in avoiding 
radio-loud sources. Are these systems related either 
physically or in an evolutionary sense to the BAL outflows? A 
physical connection is suggested by the several known cases where 
narrow-line intrinsic absorbers appear in the same spectrum 
as conventional BALs (e.g. Barlow \etal 1994; Wampler \etal 1995). 
There is evidence that some of the narrow-line \zaz\ absorbers 
in radio-loud QSOs are also intrinsic 
(Barlow \& Sargent 1997; Aldcroft \etal in this volume). 
How are those systems related to the BALs and intrinsic narrow-line 
systems in radio-quiet QSOs? Also, how are any of the intrinsic QSO 
absorbers related to the \zaz\ systems in Seyfert 1 galaxies 
(see Crenshaw \etal this volume)? A better understanding will 
require more observations to test whether narrow-line intrinsic 
systems are characteristically different in QSOs/active galaxies 
with different redshifts, luminosities and radio properties. 

\acknowledgments
This work was supported by NASA grants NAG 5-1630 and NAG 5-3234.

\end{document}